\documentclass[aps,pra,twocolumn,amsmath,amssymb,footinbib,showpacs]{revtex4-1}
\usepackage[english]{babel}
\usepackage{latexsym}
\usepackage{graphics}
\usepackage{subfigure}
\usepackage{epsfig}
\usepackage{color}
\usepackage{hyperref}
\hypersetup{
colorlinks=true,
citecolor=blue,
linkcolor=red,
urlcolor=black
}

\newcommand{\epsfboxmod}[1]{\epsfbox{#1}}

\newcommand{\infig}[2]{\begin{center}
                                    \mbox{ \epsfxsize #1 \epsfboxmod{#2}}
                                      \vspace{-0.8cm}
                                    \end{center}}

\renewcommand{\i}{\textrm{i}}
\newcommand{\ie}{{i.e.}}

\newcommand{\ch}{\textrm{cosh}}
\newcommand{\sh}{\textrm{sinh}}
\newcommand{\ftheta}[1]{#1_{\oplus}}

\newcommand{\av}[1]{\overline{#1}}
\newcommand{\VT}{V_\textrm{ho}}
\newcommand{\Vopt}{V}
\newcommand{\Vr}{V_\textrm{\tiny R}}
\newcommand{\sigmar}{\sigma_\textrm{\tiny R}}

\newcommand{\SPf}{A}
\newcommand{\green}{G}
\newcommand{\selfE}{\Sigma}

\newcommand{\DoS}{\mathcal{N}}
\newcommand{\dens}{n}
\newcommand{\rolavdens}{\tilde{\dens}}
\newcommand{\tr}{\textrm{Tr}}
\newcommand{\densmat}{\hat{\rho}}

\newcommand{\opdens}{\hat{\dens}}

\newcommand{\LTF}{L_\textrm{\tiny TF}}
\newcommand{\xiini}{\xi_\textrm{\tiny in}}
\newcommand{\pmax}{p_{\textrm{m}}}
\newcommand{\Ec}{E_{\textrm{c}}}

\newcommand{\ti}{t_{\textrm{i}}}
\newcommand{\zi}{z_{\textrm{i}}}
\newcommand{\psii}{\psi_{\textrm{i}}}
\newcommand{\densmati}{\densmat (t_{\textrm{i}})}
\newcommand{\densi}{\dens_{\textrm{i}}}
\newcommand{\thetai}{\theta_{\textrm{i}}}
\newcommand{\SCdistri}{W_{\textrm{i}}}
\newcommand{\PEdistri}{f_\textrm{i}}

\newcommand{\lyap}{\gamma}

\newcommand{\pE}{p_{\tiny E}}
\newcommand{\lyapE}[1]{\lyap{(#1)}}

\begin{document}

\title{Localization of a matter wave packet in a disordered potential}

\author{M.~Piraud$^{1}$}
\author{P.~Lugan$^{1,2}$}
\author{P.~Bouyer$^{1}$}
\author{A.~Aspect$^{1}$}
\author{L.~Sanchez-Palencia$^{1}$}
\affiliation{
$^1$Laboratoire Charles Fabry de l'Institut d'Optique,
CNRS and Univ.~Paris-Sud,
Campus Polytechnique,
RD 128,
F-91127 Palaiseau cedex, France \\
$^2$Physikalisches Institut, Albert-Ludwigs-Universit\"at, Hermann-Herder-Str. 3, D-79104 Freiburg, Germany}

\date{\today}

\begin{abstract}
We theoretically study the Anderson localization of a matter wave packet
in a one-dimensional disordered potential.
We develop an analytical model which includes the initial phase-space density of the matter wave
and the spectral broadening induced by the disorder.
Our approach predicts a behavior of the localized density profile
significantly more complex than a simple exponential decay.
These results are confirmed by
large-scale and long-time numerical calculations.
They shed new light on recent experiments with ultracold atoms
and may impact their analysis.
\end{abstract}

\maketitle

The field of Anderson localization (AL) is attracting considerable attention
and produced landmark results in the recent years~\cite{phystoday2009}.
In this respect, quantum gases stimulate intensive
experimental and theoretical
research.
On one hand, they offer
unprecedented control of their parameters,
and novel measurement tools~\cite{lsp2010},
which, for instance, paved the way to
the direct observation of one-dimensional (1D) AL of matter waves~\cite{billy2008,roati2008}.
On the other hand, they sustain original effects,
which require special analysis in its own right~\cite{lsp2007,EffMobEdge,pezze2010}.

In weak disorder, AL is due to the interference of waves
multiply scattered from random defects,
which results in absence of diffusion and
spatially localized wavefunctions~\cite{lee1985}.
The paradigmatic signature of AL is obtained from the average logarithm of the transmission of a wave
of energy $E$ through a disordered region.
In 1D, it is a self-averaging quantity characterized by the exponential decay
$\av{\ln \vert\psi_E (z)\vert^2} \simeq -2\lyapE{E} \vert z \vert$,
with $\lyapE{E}$ the Lyapunov exponent (inverse localization length)~\cite{lifshits1988}.
The localization of a wave packet is a more complicated issue
as it is determined by the superposition of many energy components.
Since the latter cannot be separated from each other,
the relevant quantity is rather the average of the localized density profile,
$\dens(z)=\av{\vert \psi (z)\vert^2}$, which is not self-averaging~\cite{lifshits1988}.
Moreover, each energy component localizes exponentially with its own localization length,
and the superposition of all their contributions can lead to
non-exponentially decaying density profiles~\cite{lsp2007,skipetrov2008}.

Localization of wave packets is for instance relevant to experiments
where a Bose-Einstein condensate (BEC)
propagates in a disordered potential~\cite{billy2008}.
The situation is modeled by the following scenario~\cite{lsp2007,shapiro2007}:
A non-disordered, interacting BEC (with initial healing length $\xiini$)
is first released from a trap. It expands in free space and
its initial interaction energy is converted into kinetic energy.
At a given time $\ti$,
a speckle potential (with correlation length $\sigmar$)
is switched on and the interactions off.
This creates a wave packet with a broad energy distribution.
The energy components are then independent and
eventually localize exponentially in the disordered potential,
which results in the localization of the matter wave.
For $\xiini > \sigmar$,
recent experiments report an exponential decay of
the density profile, in fair agreement with the prediction
of the above scenario~\cite{billy2008,hulet}.
The data however suggest deviations from
exponential decay in the wings,
the origin of which remains to be elucidated.

Here, we revisit the theoretical model for AL
of matter wave packets in 1D disorder.
Beyond previous models, our approach allows us to include
(i)~the phase-space density of the matter wave at time $\ti$, and
(ii)~the spectral broadening induced by the disorder.
We show that these ingredients significantly affect
the predicted density profile of the localized matter wave
at both short and long distances.
It predicts a complex behavior of the density profile,
which significantly deviate from pure exponential decay.
Our results are confirmed by
large-scale and long-time numerical calculations.
They shed new light on the AL of matter wave packets,
in particular on the experiments of Refs.~\cite{billy2008,hulet}.

We consider a 1D matter wave subjected to a harmonic trap and
a disordered potential, with repulsive short-range interactions.
In the weakly interacting regime
(\ie\ for large enough 1D density, $n \gg mg/\hbar^2$,
where $m$ is the atomic mass and $g$ is the coupling parameter),
its dynamics is governed by
the Gross-Pitaevskii equation,
\begin{equation}
i\hbar \partial_t \psi
= \left[
-(\hbar^2/2m) \partial_z^2
+ \VT(z)
+ \Vopt (z)
+ g|\psi|^2 -\mu
  \right] \psi, 
\label{eq:GPE} 
\end{equation}
where $\mu$ is the chemical potential,
the wavefunction is normalized to the total number of atoms ($\int \textrm{d}z |\psi|^2 = N$),
$\VT(z)=m\omega^2z^2/2$ is the trapping potential,
and $V(z)$ is the disordered potential.
The latter is assumed to be stationary with a null ensemble average, $\av{V}=0$.
It is characterized by the correlation function
$C(z)=\av{V(z^\prime) V(z^\prime+z)}$.
Hereafter, the quantities $\Vr$ and $\sigmar$
denote the amplitude and correlation length of the disorder
(see below for precise definitions).
We define the healing length of the trapped BEC
by $\xiini \equiv \hbar/\sqrt{4m\mu}$.
In the following, we study the average density profile:
$\dens (z,t) = \av{\tr [\densmat(t)\opdens(z)]}$,
with $\densmat$, the one-body density matrix
and $\opdens(z)=\delta(z-\hat{z})$, the spatial density operator.

Following the scenario of Refs.~\cite{lsp2007,shapiro2007},
an interacting BEC is first produced in the harmonic trap
and in the absence of disorder.
For interactions strong enough that $n \gg \hbar\omega/g$ [Thomas-Fermi regime (TF)],
the phase is uniform and the density profile is a truncated inverted parabola,
$\dens_0(z) = (\mu/g) \ftheta{[1-(z/\LTF)^2]}$,
where $\LTF = \sqrt{2\mu/m\omega^2}$ and
$\ftheta{[f(z)]}=f(z)$ for $f(z)>0$ and $0$ otherwise.
Then, an expanding matter wave is produced by switching off the trap
($\VT \rightarrow 0$) at time $t=0$.
We assume that,
in the first expansion stage ($0\leq t\leq \ti$),
the disordered potential is still off,
so that the density matrix at time $\ti$ is pure state:
$\densmati = \vert \psii\rangle\langle \psii \vert$,
where $\psii (z)=\textrm{e}^{\i\thetai (z)}\sqrt{\densi (z)}$
is determined by the integration of Eq.~(\ref{eq:GPE})
with $V \equiv 0$.
The solution reads~\cite{scaling}
\begin{equation}
\densi (z) = \dens_0 \big(z/b(\ti)\big)/b(\ti)
~~\textrm{and}~~
\thetai(z) = m z^2 \dot{b}(\ti) / 2\hbar b(\ti),
\label{eq:scaling}
\end{equation}
where the scaling parameter is the unique solution of
$\sqrt{b(t) [b(t)-1]} + \ln [\sqrt{b(t)}+\sqrt{b(t)-1}] = \sqrt{2} \omega t$
\cite{clement2005}.
For $t \gg 1/\omega$, we have
$b(t) \simeq \sqrt{2}\omega t + ({1}/{2})[1-\ln (4\sqrt{2}\omega t)]$.

The second expansion stage ($t > \ti$) starts when
the disorder is suddenly switched on and the interactions off
($V \not\equiv 0$ and $g \rightarrow 0$).
Then, Eq.~(\ref{eq:GPE}) reduces to the linear Schr\"odinger equation
of Hamiltonian $\hat{H} = -\hbar^2\partial_z^2/2m + V(z)$
with the initial state given by Eq.~(\ref{eq:scaling}) at time $\ti$.
Turning to the Heisenberg picture, we get:
\begin{equation}
\dens (z,t) = \av{\tr [\densmati \opdens(z, t-\ti)]}.
\label{eq:dens0}
\end{equation}
We now treat the \textit{initial} state of the second expansion stage
semiclassically and apply the substitution
$\densmati \rightarrow \int dz dE\
\delta (z-\hat{z}) \DoS (E)^{-1} \PEdistri (z,E) \delta (E-\hat{H})$,
where $\DoS (E)$ is the density of states per unit length associated
to the Hamiltonian $\hat{H}$ above
and $\PEdistri (z,E)$ represents the probability density to find an atom
at position $z$ with energy $E$~\cite{note:Wigner}.
Inspection of Eq.~(\ref{eq:scaling}) legitimizes the semi-classical approximation:
The initial state is characterized by the velocity field
$v_\textrm{i}(z) \equiv (\hbar/m)\partial_z\thetai = z\dot{b}(\ti)/b(\ti)$,
associated to the local de Broglie wavelength
$\lambda_\textrm{dB}(z) \equiv \hbar/mv_\textrm{i}(z) \sim \hbar\ti/mz$
for $\ti \gg 1/\omega$, and 
we find $\lambda_\textrm{dB}(z) \ll \densi/\vert\partial_z\densi\vert$,
except in a small region of width $\Delta z \sim \xiini$ near the edges of the BEC.
For $\ti \gg 1/\omega$, we can then use the
phase-space distribution
$\SCdistri(z,p) \simeq \densi ( z ) \times \delta \big(p-mv_\textrm{i}(z)\big)$,
\ie\
\begin{equation}
\SCdistri(z,p)
\simeq
\mathcal{D}_{\textrm{i}} (p) \times \delta \big[z-\big(b(\ti)/m\dot{b}(\ti)\big)p\big]
\label{eq:SCdistr1}
\end{equation}
where
$\mathcal{D}_{\textrm{i}} (p) = \big({3N}/{4\pmax (\ti)}\big)\ \ftheta{[1 - ( p / \pmax (\ti) )^2 ]}$
is the momentum distribution,
with $\pmax(\ti) \equiv (\hbar/\xiini)(\dot{b}(\ti)/\sqrt{2}\omega)$
\cite{note:pmaxinf}.
Averaging over the disorder,
we can then write
$\av{\PEdistri (z,E)} \simeq \int dp\ \SCdistri (z,p) \SPf (p,E)$,
where
$\SPf (p,E) = -\textrm{Im} \big\langle p \big\vert \av{\green} (E) \big\vert p \big\rangle/\pi$
is the spectral function,
which represents the probability density that a particle in the state of momentum $p$,
$\vert p \rangle$, has energy $E$
in the
disorder.
Here, $\green (E)= [E-\hat{H} + \i 0^+]^{-1}$ is the retarded Green operator
associated to Hamiltonian $\hat{H}$ at energy $E$~\cite{skipetrov2008}.
In order to evaluate the BEC density,
we finally insert these formulas
into the
rhs term of 
Eq.~(\ref{eq:dens0})~\cite{note:DisoAv}, which yields
\begin{equation}
n (z,t) =
\int\!\! d\zi dE \int\!\! dp\ \SCdistri (\zi,p) \SPf (p,E)
P(z-\zi, t-\ti \vert E)
\label{eq:denstot}
\end{equation}
with
$P(z \! - \! \zi, \tau \vert E) \equiv \DoS (E)^{-1}
\av{\tr [ \delta (E \! - \! \hat{H}) \opdens (z,\tau) \delta (\zi \! - \! \hat{z}) ]}$.
The quantity $P(z-\zi, t-\ti \vert E)$ is interpreted as the probability of quantum diffusion,
that is the probability density to find in $z$ at time $t$,
a particle of energy $E$ that was located in $\zi$ at time $\ti$ \cite{berezinskii1974}.

Equation~(\ref{eq:denstot})
allows us
to determine
the density profile of the matter wave packet.
Our approach goes beyond that of Ref.~\cite{lsp2007}.
It allows us to take into account
(i)~the initial position distribution and
(ii)~the spectral broadening $\SPf (p,E)$
of a particle of momentum $p$ in the disordered potential~\cite{note:CFlsp2007}.
We will show below that both play a significant role in the
localization process.

In order to calculate the average spectral function, we solve the Dyson equation,
$\av{\green} = \green_0 + \green_0 \selfE \av{\green}$,
where $\green_0 (E) = [E-\hat{p}^2/2m + \i 0^+]^{-1}$ is the free Green operator
and $\selfE (E) = \selfE^\prime (E) + \i\selfE^{\prime\prime} (E)$
is the self-energy, both in the retarded form.
For $\Vr^2 \ll E^{3/2} \Ec^{1/2}$
where $\Ec \equiv \hbar^2/2m\sigmar^2$
[\ie\ $\lyapE{E} \ll \pE/\hbar$],
we find~\cite{rammer2004}
\begin{equation}
\SPf (p,E) =
\frac{(-1/\pi)\ \selfE^{\prime\prime}(E,p)}{\big(E-p^2/2m-\selfE^{\prime}(E,p)\big)^2 + \selfE^{\prime\prime}(E,p)^2}
\label{eq:spectral}
\end{equation}
with
$\selfE (E,p) \simeq \big\langle p \big\vert \av{V \green_0(E) V} \big\vert p \big\rangle$.
Performing the integration, we get the explicit formula
\begin{equation}
\selfE^{\prime\prime} (E,p) \simeq
-({m}/{2\hbar \pE})
\big\{
\tilde{C} (\pE - p)
+ \tilde{C} (\pE + p)
\big\}
\label{eq:invMFtime}
\end{equation}
where $\tilde{C}(p) \equiv \int dz\ C (z) \exp (-\i p z /\hbar)$
is the Fourier transform of the correlation function,
and $\pE \equiv \sqrt{2mE}$ is the momentum associated to energy $E$ in free space.
The real-part of the self-energy,
$\selfE^{\prime}(E,p) = \int \frac{dq}{2\pi\hbar} \tilde{C}(q - p)
\times\textrm{PV}\big(\frac{1}{E-q^2/2m}\big)$
with $\textrm{PV}$ the Cauchy principal value,
turns out to be negligible and we disregard it in the remainder of this work.

In order to calculate the probability of quantum diffusion,
we rely on the diagrammatic method developed in Refs.~\cite{berezinskii1974,gogolin1976}.
In the weak disorder limit ($\lyapE{E} \ll \sigmar^{-1}, \pE/\hbar$),
it provides the infinite-time limit~\cite{note:misprint}
\begin{eqnarray}
P_\infty(z \vert E)
& = &\frac{\pi^2 \lyapE{E}}{8} \int_0^\infty\! \textrm{d}u\ 
u\ \sh (\pi u) \left[ \frac{1+u^2}{1+\ch (\pi u)} \right]^2
\nonumber \\
& & \times \textrm{exp}\left\{- (1+u^2) \lyapE{E} |z| / 2 \right\}
\label{eq:gogolin}
\end{eqnarray}
with the Lyapunov exponent
\begin{equation}
\lyapE{E} \simeq
({m^2}/{2\hbar^2 \pE^2})\
\tilde{C}(2\pE).
\label{eq:lyapunov}
\end{equation}

We now compare the predictions of our
analytical model to the results of numerical calculations.
In order to integrate Eq.~(\ref{eq:GPE}),
we use a Crank-Nicolson algorithm,
with
space step
$\Delta z = 0.1\sigmar$
and time step
$\Delta t = 0.05 \hbar/\Ec$,
in a box with Dirichlet boundary conditions~\cite{note:num}.
The disorder is a 1D speckle potential,
which can be written in the form  $V(z)=\Vr\times\{v(z)-\av{v}\}$,
where $v(z) \geq 0$ represents the speckle intensity pattern
and the sign of $\Vr$ depends on the detuning of the laser light
with respect to the atomic resonance:
$\Vr>0$ for blue detuning, and $\Vr<0$ for red detuning~\cite{goodman2007,clement2006}.
We use parameters close to those of Ref.~\cite{billy2008}
and both blue and red detunings
($\Vr = \pm 0.0325\Ec$).
In both cases, the correlation function reads
\begin{equation}
\tilde{C}(p) = \pi\Vr^2\sigmar \ftheta{[1 - \vert p \vert \sigmar/2\hbar]},
\label{eq:Cspeckle}
\end{equation}
in Fourier space.
Our numerical calculations differ from those of Ref.~\cite{lsp2007}
in that we use significantly larger boxes and longer times.
Moreover, we consider here exactly the above scenario
where the disorder is switched on and the interactions off at a time $\ti \gg 1/\omega$,
while in the numerics of Ref.~\cite{lsp2007}, both disorder and interactions
were on during the whole expansion.
Due to the cutoffs of $\tilde{C}(p)$ at $p=\pm 2\hbar/\sigmar$,
the quantities $\lyapE{E}$ and $P_\infty(z \vert E)$ vanish for $E>\Ec$
and it is useful to distinguish two cases
for the analysis of the density profiles.

\begin{figure}[t!]
\infig{27em}{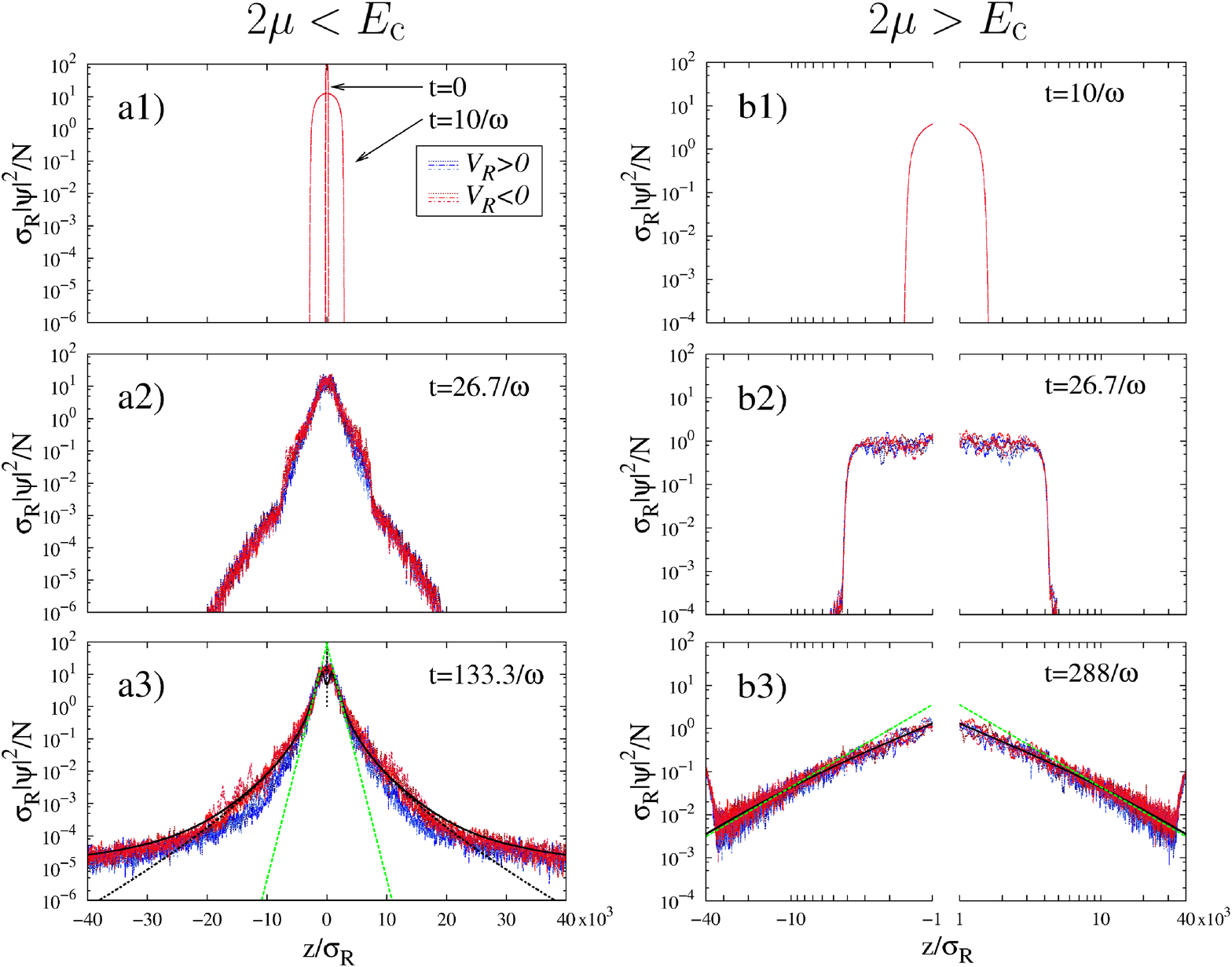}
\caption{
\label{fig:model}
(color online)
Time evolution of the density profile of a matter wave expanding
in 1D speckle potentials for $2\mu < \Ec$ (left panel) and $2\mu > \Ec$ (right panel).
Shown are the running averages,
$\rolavdens (z) = \int_{-l/2}^{+l/2} \frac{dx}{l} \dens (z \! + \! x)$
with $l = 100\sigmar$, of numerical data,
for blue- and red-detuned speckle potentials (three realizations each)~\cite{note:num}.
Here, we use
  $\omega = 2 \times 10^{-2}\mu/\hbar$
and
  $\Vr = \pm 0.0325\Ec$.
Left panel [(a1)-(a3)]: semi-logarithmic scale for $\xiini = 1.5\sigmar$ ($2\mu \simeq 0.44\Ec$).
The solid black line shows a fit of the full Eq.~(\ref{eq:denstot})
and
the dotted black line to Eq.~(\ref{eq:denstot})
with $\SPf (p,E) \rightarrow \delta (E \! - \! p^2/2m)$,
both with a multiplying factor as the only fitting parameter.
The dashed green line is a fit of $\ln [n(z)] = A -2\lyapE{2\mu}\vert z\vert$,
with $A$ as the fitting parameter.
Right panel [(b1)-(b3)]: log-log scale for $\xiini = 0.83\sigmar$ ($2\mu \simeq 1.44\Ec$).
The solid black line shows the full Eq.~(\ref{eq:denstot})
and the dashed green line is a fit of $n(z) = A / \vert z \vert^\beta$
with $A$ and $\beta$ as the fitting parameters.
}
\end{figure}

\paragraph*{Case $2\mu < \Ec$ (\ie\ $\xiini > \sigmar$)~-~}
The left panel of Fig.~\ref{fig:model} shows the
time evolution of the density profile in semi-logarithmic scale 
for $\xiini > \sigmar$.
For $ t< \ti \equiv 10/\omega$ [Fig.~\ref{fig:model}(a1)],
the matter wave expands in free space,
with the shape of a truncated inverted parabola of increasing size,
according to Eq.~(\ref{eq:scaling}).
When the disorder is switched on and the interactions off,
the matter wave continues expanding and develops long wings
[Fig.~\ref{fig:model}(a2)].
In the long-time limit [Fig.~\ref{fig:model}(a3)],
the density profile converges to a stationary shape,
hence demonstrating AL.
The localized density profile is in fair agreement with the theoretical prediction
based on Eq.~(\ref{eq:denstot}),
using Eqs.~(\ref{eq:SCdistr1}), (\ref{eq:spectral}) and (\ref{eq:gogolin})
for the correlation function~(\ref{eq:Cspeckle}),
with a global multiplying factor as the only fitting parameter
[see the solid black line in Fig.~\ref{fig:model}(a3)].
This holds over the full space, except very close to the center.
There, for the chosen parameters, Eq.~(\ref{eq:denstot}) predicts a nonphysical dip
due to the overestimation of the Lyapunov exponents and the spectral broadenings
at low energies in the lowest-order perturbation theory
used to derive Eqs.~(\ref{eq:spectral}) and (\ref{eq:gogolin}).
This dip affects the balance between the center and the wings in the 
normalization of the wavefunction, which justifies the multiplying factor 
to correctly fit the wings.
These results validate the localization model of the matter wave.

Let us now discuss the density profile in more detail,
and accordingly examine the impact of the various terms in Eq.~(\ref{eq:denstot}).
For
$\vert z \vert \lesssim b(\ti)\LTF$,
the stationary density profile is mainly determined by particles originating
from the BEC at time $\ti$ that propagate over very short distances in the disordered potential.
Using the full phase-space distribution $\SCdistri (z,p)$
is then necessary to account for the central feature of the density profile.
For $\vert z \vert \gtrsim b(\ti)\LTF$ only,
we can neglect the initial density distribution
and rely on the approximation $\SCdistri(z,p) \rightarrow \mathcal{D}(p) \! \times \! \delta(z)$
in Eq.~(\ref{eq:denstot}).
For
$b(\ti)\LTF \lesssim \vert z \vert \lesssim 1/2\lyapE{2\mu}$,
we find that the density profile shows an essentially exponential decay
of rate approximately equal to $2\lyapE{2\mu}$
[see the dashed green line in Fig.~\ref{fig:model}(a3)].
This is consistent with experimental observations~\cite{billy2008}.
For longer distances however,
the logarithmic derivative of the density continuously decreases
in modulus.
Neglecting the spectral broadening induced by the disorder~\cite{lsp2007,shapiro2007},
$\SPf (p,E) \rightarrow \delta (E \! - \! p^2/2m)$,
we are able to reproduce the numerical results
over about five decades [see the dotted black line in Fig.~\ref{fig:model}(a3)].
This approximation cuts all components with $E>2\mu$
and predicts a long-distance exponential decay of rate
$\lyapE{2\mu}/2$ \cite{lsp2007,note:misprint}.
This behavior can be understood on the basis of
the probability of quantum diffusion~(\ref{eq:gogolin}), which continuously interpolates
from ${d\ln P_\infty (z \vert E)}/{dz} \simeq -2\lyapE{E}$
for $\vert z \vert \ll 1/2\lyapE{E}$
to ${d\ln P_\infty (z \vert E)}/{dz} \simeq -\lyapE{E}/2$
for $\vert z \vert \gg 2/\lyapE{E}$~\cite{gogolin1976}.
For $\vert z \vert \gg 2/\lyapE{2\mu}$,
the numerics show significant deviation from exponential decay,
owing to the Lorentzian-like form of the spectral function~(\ref{eq:spectral})
which populates components with $E>2\mu$.
Then, taking into account the full spectral function,
Eq.~(\ref{eq:denstot}) fits the numerics well
[see the solid black line in Fig.~\ref{fig:model}(a3)].
Finally, note that our model relies on the Born approximation which is not sufficient
to account for components with $E>\Ec$~\cite{EffMobEdge}.
To do so, it would be necessary to include arbitrary high-order terms
at arbitrary large distances.
It however appears irrelevant in the space window used for the numerics.

\paragraph*{Case $2\mu > \Ec$ (\ie\ $\xiini < \sigmar$)~-~}
The right panel of Fig.~\ref{fig:model} shows the counterpart of the left panel
for $\xiini < \sigmar$ and in log-log scale.
In this regime too, the complete model of Eq.~(\ref{eq:denstot}) reproduces
well the numerical results over the full space (except very close to the center),
with a multiplying factor as the only fitting parameter
[see the solid black line in Fig.~\ref{fig:model}(b3)].
In 1D speckle potentials, the correlation function
provides a high-momentum cutoff which strongly suppresses backscattering of matter waves
with momentum $p > \hbar/\sigmar$~\cite{lsp2007,EffMobEdge}.
For $E>\Ec$, the Lyapunov exponent, calculated
in the Born approximation then vanishes [see Eq.~(\ref{eq:lyapunov})],
and the determination of $P_\infty (z \vert E)$
would require an extension of the formalism of Refs.~\cite{berezinskii1974,gogolin1976}
by at least two orders in perturbation theory.
Using the results of Refs.~\cite{EffMobEdge} based on the phase-formalism approach,
we estimate that, for our parameters, the Lyapunov exponent
drops by about two orders of magnitude around $E \simeq \Ec$
and we neglect localization of waves with $E > \Ec$.
Since $\Ec<2\mu$, the spectral broadening
has little importance here
and we can safely rely on
the approximation $\SPf (p,E) \rightarrow \delta (E \! - \! p^2/2m)$.
The above model then predicts algebraic localization,
$\dens (z) \propto 1/\vert z \vert^2$~\cite{lsp2007}.
Fitting an algebraic function, $A/\vert z \vert^\beta$ with $A$ and $\beta$ as fitting parameters,
to the numerical data of three different realizations of blue- and red-detuned
speckle potentials
in the intervals $[-30,-10]$ and $[+10,+30]$ independently,
we find $\beta \simeq 1.91 \pm 0.22$.
This is in fair agreement with the analytical prediction
(within the error bars)
and was observed in Ref.~\cite{billy2008}.

In summary, we have developed a theoretical model for
the AL of a matter wave packet with initial healing length $\xiini$
in a 1D speckle potential with correlation length $\sigmar$.
It extends previous approaches by including
(i)~the initial phase-space density of the matter wave, and
(ii)~the spectral broadening induced by the disorder.
We have shown that these ingredients affect
the localized density profiles,
which significantly deviate from a pure exponential decay.
For $2\mu < \Ec$,
we found that $n(z)$ essentially shows an exponential decay
of rate $2\lyapE{2\mu}$ at short distance, in accordance with experimental observations~\cite{billy2008}.
For larger distance, $n(z)$ crosses over to an exponential decay of rate $\lyapE{2\mu}/2$
and then deviates from exponential decay due to the disorder-induced spectral broadening.
This may explain the very large distance behavior of experimental data~\cite{billy2008,hulet}.
For $2\mu > \Ec$,
we found algebraic localization, $n(z) \propto 1/\vert z \vert^2$,
as observed in Ref.~\cite{billy2008}.

In the future, it would be interesting to extend the present approach toward two directions.
First, our analysis relies on the calculation of
the probability of quantum diffusion, $P_\infty (z \vert E)$, to lowest order,
which is valid only below the effective mobility edge at $E=\Ec$~\cite{EffMobEdge}.
Extending the diagrammatic method of Refs.~\cite{berezinskii1974,gogolin1976}
to higher orders would allow one to incorporate the components of energy
$E>\Ec$.
In addition, it may explain the slight difference in the localized density profiles
found for blue- and red-detuned speckle potentials [see Fig.~\ref{fig:model}(a3)].
Second, although ultracold atoms allow for an exact realization of the above scenario
using time-dependent control of optical disorder
and of interactions via Feshbach resonance techniques,
recent experiments have followed a slightly different scheme
where the BEC is created already in the presence of the disorder
and the interactions are not switched off~\cite{billy2008,hulet}.
Extending our model to this case would require one to include
(i)~the effect of the disorder at $t \lesssim 1/\omega$,
which can significantly modify the relevant phase-space density $\SCdistri (z,p)$
and
(ii)~the effect of interactions in the probability of quantum diffusion $P_\infty (z \vert E)$.
Whether and how interactions destroy localization in this scheme is
still a very debated subject~\cite{nonlinearloc}.

We thank Peter Schlagheck, Luca Pezz\'e, Dominique Delande, Georgy Shlyapnikov and Randy Hulet
for stimulating discussions. 
This research was supported by
the European Research Council (FP7/2007-2013 Grant Agreement No.\ 256294),
Agence Nationale de la Recherche (ANR-08-blan-0016-01),
Minist\`ere de l'Enseignement Sup\'erieur et de la Recherche,
Triangle de la Physique and
Institut Francilien de Recherche sur les Atomes Froids (IFRAF).
We acknowledge GMPCS high performance computing facilities of the LUMAT federation.



\begin{references}
\bibitem{note:Wigner}
We may use alternative semi-classical approaches.
For instance, the Wigner distribution provides similar results.
In fact, Eq.~(\ref{eq:SCdistr1}) accurately approximates the latter.
\bibitem{note:pmaxinf}
In the long-time limit ($\ti \rightarrow \infty$),
we find
$\pmax \rightarrow \hbar/\xiini$.
\bibitem{note:DisoAv}
This approach neglects possible correlations in disorder-averaging
of the initial state and the propagation kernel.
For weak disorder, it is reasonable to assume they are small,
which is supported by the good agreement between Eq.~(\ref{eq:denstot})
and numerical calculations (see text).
\bibitem{note:CFlsp2007}
We recover the formulation of Ref.~\cite{lsp2007} by using
the substitutions
$\SCdistri(z,p) \! \rightarrow \! \mathcal{D}(p) \! \times \! \delta(z)$,
which holds at distances exceeding the initial size of the gas,
and $\SPf (p,E) \! \rightarrow \! \delta (E \! - \! p^2/2m)$, which is the free spectral function.
\bibitem{note:num}
Only the central half of the box used in the numerics
is plotted in Fig.~\ref{fig:model}.
We have checked that the boundary conditions do not significantly
alter the results
in the represented space window by using boxes of different sizes.
\end{references}
\end{document}